# Distribution of Relaxation Times analysis of evolution of Oxygen Reduction Pathways for ionic conductor infiltration on MIEC cathode

Rakhi Saha[1], Preethi Sudarsan[2], Manju Kumari[1], Hee Jung Park[2] Abdelkrim Mekki[3,4], Khalil Harrabi[4,5], Somaditya Sen[1]*

[1]*Department of Physics, Indian Institute of Technology Indore, Indore, 453552, India*
[2]*Department of Materials Science and Engineering and Hydrogen Research Center, Dankook University, Cheonan 31116, Republic of Korea*
[3] *Department of Physics, King Fahd University of Petroleum & Minerals Dhahran, 31261, Saudi Arabia*
[4]*Interdiciplinary Research Center (IRC) for Advanced Material, King Fahd University of Petroleum & Minerals, Dhahran 31261, Saudi Arabia*
[5]*Interdisciplinary Research Center (RC) for Intelligent Secure Systems, King Fahd University of Petroleum & Minerals, Dhahran 31261, Saudi Arabia*

**Corresponding author: sens@iiti.ac.in*

**Keywords:** Distribution of Relaxation Times, Oxygen Reduction Reaction, Cathode infiltration, Polarization resistance, Impedance spectroscopy, Ionic-electronic conductivity

## Abstract:

Solid oxide fuel cells (SOFCs) are promising electrochemical energy-conversion devices; however, the sluggish cathodic oxygen reduction reaction (ORR) remains a major limitation for intermediate-temperature operation. ORR subprocesses can be modified by infiltrating ionic $Sm_{0.2}Ce_{0.8}O_{2-\delta}$ (SDC) on mixed ionic–electronic $SrFe_{0.9}Ti_{0.1}O_{3-\delta}$ (STF). However, the processes are indistinguishable in most cases and poorly understood using conventional equivalent circuits. Distribution of Relaxation Times (DRT) analysis distinguishes these processes to identify the dynamics with temperature and surface reconstruction. Such analysis is being reported, revealing the connection between increase of active sites, temperature, and polarization resistance ($R_P$). SDC infiltration preferentially accelerates oxygen surface exchange and activation processes over the relatively high-frequency charge transfer process related to cathode surface at elevated temperatures for the infiltrated cells. $R_P$ was reduced substantially with the systematic redistribution of each process to as low as 0.04 Ω·cm² at 800 °C. This work underlines the deconvolution of the processes using DRT as a tool, and SDC infiltrated STF as a model cathode system to provide a mechanistic insight of understanding ORR kinetics and design a rationale for developing high-performance SOFC air electrodes.

## Introduction:

In the current demand of clean energy resources, Solid Oxide Fuel Cell (SOFC) are the most promising electrochemical energy conversion devices due to their high energy efficiency, fuel flexibility and reversible mode of operation.[1–4] However, their widespread commercialization is hindered by its high operational temperature (800°C-1000°C), thermal expansion co-efficient (TEC) mismatch between constituent cell components, long-term material degradation and issues related to interconnect poisoning.[5,6] Consequently, efforts have been devoted among fuel cell researchers to lower its operational temperature to intermediate temperature referred as IT-SOFC (600°C-800°C).[7–9] This is expected to improve material durability and mitigate degradation-related issues. But the high activation energy (~ 1.3 -1.88 eV) of the cathodic oxygen reduction reaction (ORR) $[\frac{1}{2}\,O_2\,(g) + 2e^-\,(cathode) \rightarrow\,O^{2-}\,(electrolyte)]$ deteriorates the cell performance with decrease in temperature.[10] ORR in SOFC is a highly complex mechanism and sensitive to electrode composition, microstructure, and operating temperature. It depends on multiple intermediate steps including gas diffusion through the porous cathode matrix, physio-adsorption followed by chemosorption, ionization and dissociation of oxygen ions, incorporation of reduced oxygen ions into oxygen vacancy ($V_O$) of cathode and finally transported to electrolyte,[11,12] and is difficult to deconvolute. Hence, addressing each component step becomes a problem. This is a bottleneck in improving the cell performance in IT-SOFCs. One of the important factors in determining the cathode performance is its polarization resistance ($R_P$) which is a collective resistance offered by all these subprocesses. Electrochemical Impedance Spectroscopy (EIS) is a widely employed in-situ technique for investigating electrode kinetics and evaluating $R_P$.[13] The EIS spectra contain the information of all individual processes each being represented by an equivalent resistor-capacitor (R-C) circuit. Hence, in SOFCs, the presence of simultaneous multiple interfaces and different types of active sites, multiple R-C combinations become necessary. A unique solution becomes difficult as processes, closely spaced in frequency, overlap and are often inseparable. This makes complex non-linear spectra (CNLS) fitting questionable.[14,15] To overcome these limitations of CNLS, Distribution of Relaxation Times (DRT) analysis has emerged as a powerful approach for resolving overlapping electrochemical processes in the relaxation-time domain.[16–18]

To enhance ORR kinetics, one needs cathode materials with improved electronic and ionic conductivity. To achieve such goals surface engineering strategies such as infiltration and composite electrode formation have been explored. [19–23] However, most studies primarily focus on performance enhancement, while the evolution of the underlying ORR kinetics remains less understood. To understand the evolution of these underlying processes a choice of a mixed ionic-electronic conductor (MIEC) with surface modification by a good ionic conductor should be an ideal model. $SrFeO_{3-\delta}$ (SFO) is a promising MIEC air-electrode (cathode) material owing to its considerable $V_O$ concentration ($\delta \approx 0.13$–$0.40$) and $Fe^{3+/4+}$ charge disproportionation.[24,25] However, the oxygen-deficient Tetragonal ($\delta \approx 0.14$) and Orthorhombic ($\delta \approx 0.27$) phases suffer from structural instability during repeated oxidation–reduction cycling.[26] To address this issue, higher-valence cations are commonly introduced to stabilize the crystal structure. 10% $Ti^{4+}$ substitution at the B-site ($SrFe_{0.9}Ti_{0.1}O_{3-\delta}$: STF) retains SFO cubic phase.[27,28] The electronic conductivity is slightly affected but is considerable due to the retainment of ample number of $Fe^{3+/4+}$ redox couples in the STF lattice. On the other hand, the cubic phase helps structural integrity by preventing $V_O$ ordering, reducing the possibility of structural transformation to a less conducting brownmillerite phase of SFO.[29] However, a reduction in $V_O$ concentration is accompanied by decreased ionic conductivity. To address this issue and improve the ionic conductivity while protecting the bulk STF structure, successive infiltrations of an ionically conducting material like Sm-doped Ceria (SDC)[30] at the STF interface is a probable mechanism of obtaining an ionic-electronic conducting composite. Active reaction sites (triple phase boundaries (TPBs)) can be controlled by the amount of infiltration. This helps in modifying the ratios of ionic to electronic conduction pathways. Moreover, the infiltration technique requires annealing at comparatively lower temperatures (600°C ~ 800°C) than conventional high temperature sintering (>1000°C) for composite materials. Such modifications in the TPBs are an ideal model to investigate ORR kinetics and use DRT as a tool to analyze the different processes in such MIEC systems. The result of infiltration is beneficial in obtaining a comparably low $R_P$ w.r.t. other SFO derived cathode materials. Note that most DRT studies on SOFC electrodes primarily focuses on qualitative process identification based on spectra evolution with temperature or operating conditions without discussing the role of individual processes with electrode modifications. This work reports a detailed understanding of how successive infiltration cycles (schematic of infiltration in figure S1) redistribute individual ORR processes and alter the dominant kinetic bottlenecks. Four different sets of infiltrated cells were prepared and the sample with N-number infiltration (IF) will be referred to in this manuscript as IFN. Progressive modification of the infiltrated SDC layers on electrode surface was characterized by X-ray Diffraction (XRD) and Field-Emission Scanning Electron Microscopy (FESEM), while the electrochemical response was evaluated using temperature-dependent (650°C – 800°C) EIS. DRT peak fitting was employed to resolve individual electrochemical processes and monitor their evolution with composition and temperature, providing a valuable guideline for a rational analysis of IT-SOFCs.

**Results and discussion:**

From the room-temperature XRD studies, cubic *(Pm-3m)* perovskite $ABO_3$, SFO-like structure was observed for the STF powder sample (Figure 1a). The sharp peaks indicate the high crystallinity of the material. Major peaks were observed at [(hkl) → 2θ (FWHM)]: (110) → 32.71 (0.165), (111) → 40.35 (0.185), (200) → 47.00 (0.215), (211) → 58.33 (0.26), (220) → 68.60 (0.47622). The Williamson-Hall analysis was performed to extract the crystallite size (97 nm) and lattice strain (0.13) of the nanoparticles. The absence of any impurity phases confirms the single-phase nature of the sample. The structural refinement with a cubic CIF file (ICSD-1528364) provided a good fit with a goodness of fit parameter ($\chi^2$) value of 1.63. The refined structural parameters are listed in Table S1, and the refinement plot is shown in Figure 1b.

The SDC powder revealed a pure single fluorite cubic phase (*Fm-3m*) (Figure S2) with major peaks were observed at [(hkl) → 2θ (FWHM)]: (111) → 28.43 (0.404), (200) → 32.95 (0.404), (220) → 47.30 (0.451), (311) → 56.12 (0.471), (222) → 58.85 (0.476), (400) → 69.14 (0.420), (331) → 76.38 (0.540), (420) → 78.76 (0.541). The Williamson-Hall analysis was performed to extract the crystallite size (21 nm) and lattice strain (0.037) of the nanoparticles. The same SDC powder was used for preparing the buffer layer.

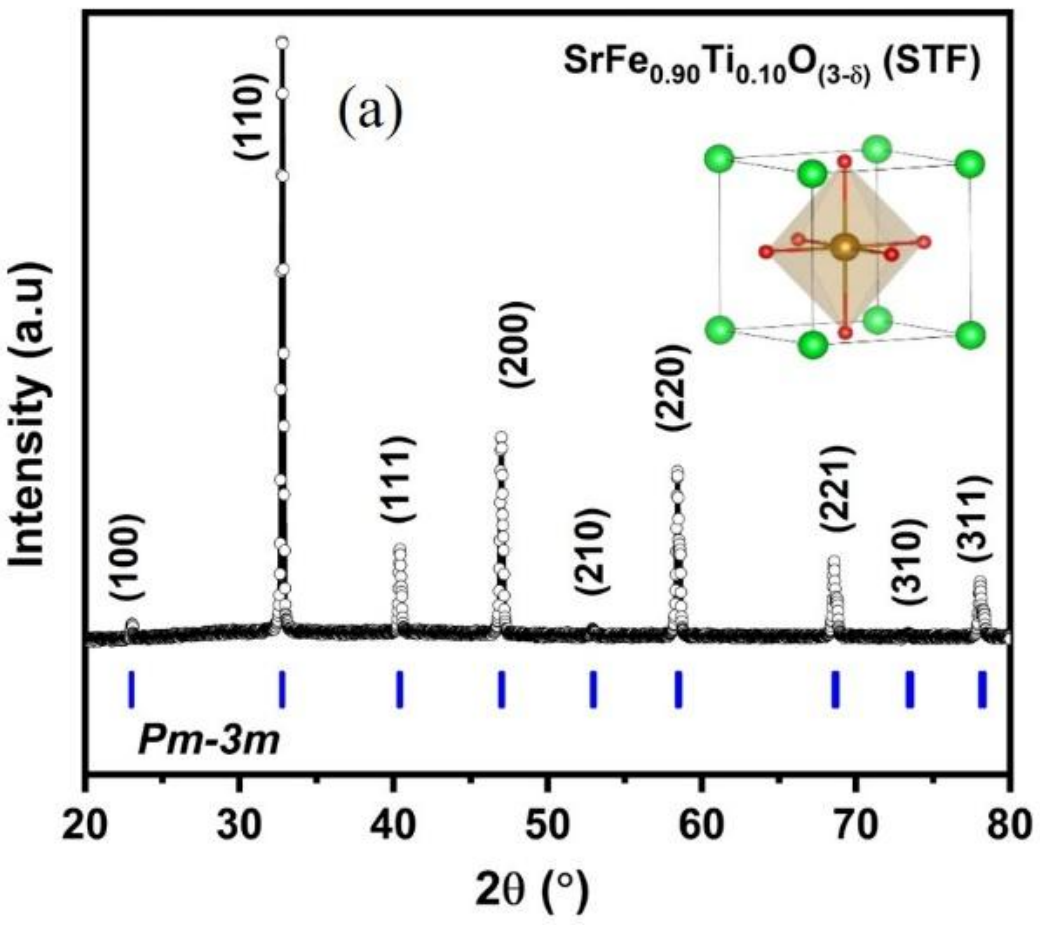

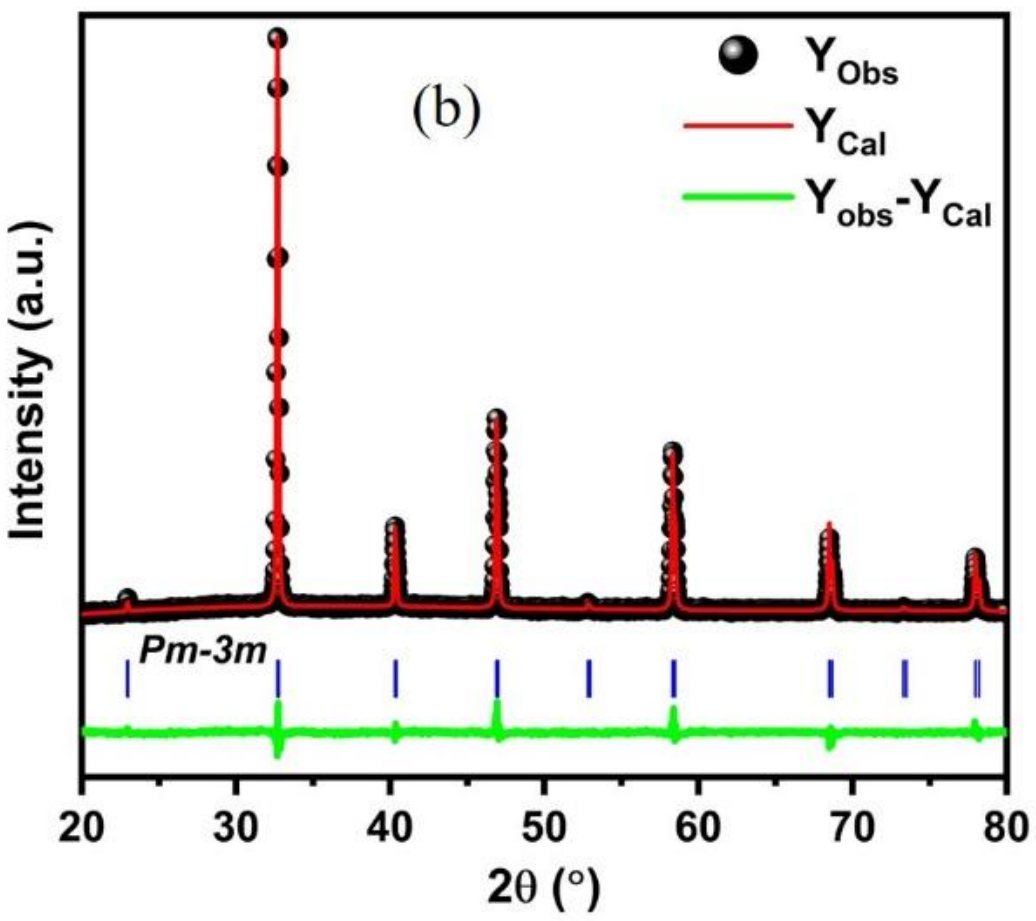

**Figure 1:** (a) XRD pattern of the calcined $SrFe_{0.90}Ti_{0.10}O_{3-\delta}$ (STF) powder with Cubic (Pm-3m) CIF without the presence of any impurity peak; (b) Rietveld refinement of STF with Pm-3m space group with observed (black dot), fitted (red line) and difference (green line)

The cross-sectional SEM images of pristine STF and IF symmetric cells after the final testing at 800°C are shown in figure 2a. The images clearly reveal the dense YSZ electrolyte layer sandwiched between the SDC buffer layers. A uniform distribution of Y and Zr was observed from elemental mapping (EDX) micrographs throughout the electrolyte region, confirming the compositional homogeneity of the dense YSZ electrolyte (Figure – 2b, c, d). The SDC buffer layer revealed the presence of proportionate amounts of Sm and Ce (Figure – 2e, f), along with uniform Fe distribution in STF cathode (Figure - 2g). Thickness values were estimated using ImageJ software: the YSZ electrolyte prepared by tape-casting method was measured to be of thickness ~ 150 – 170μm, the SDC buffer layer was < 14μm, and the screen-printed STF cathode layer was ~ 30μm composed of microparticles of size ~ 370 nm. The microstructure of the pristine STF scaffold (Figure 3a) exhibits high porosity, which is favorable for infiltration. The IF1 and IF4 showed increased density of SDC nanoparticles on the cathode backbone (Figure – 3b, c, d)).

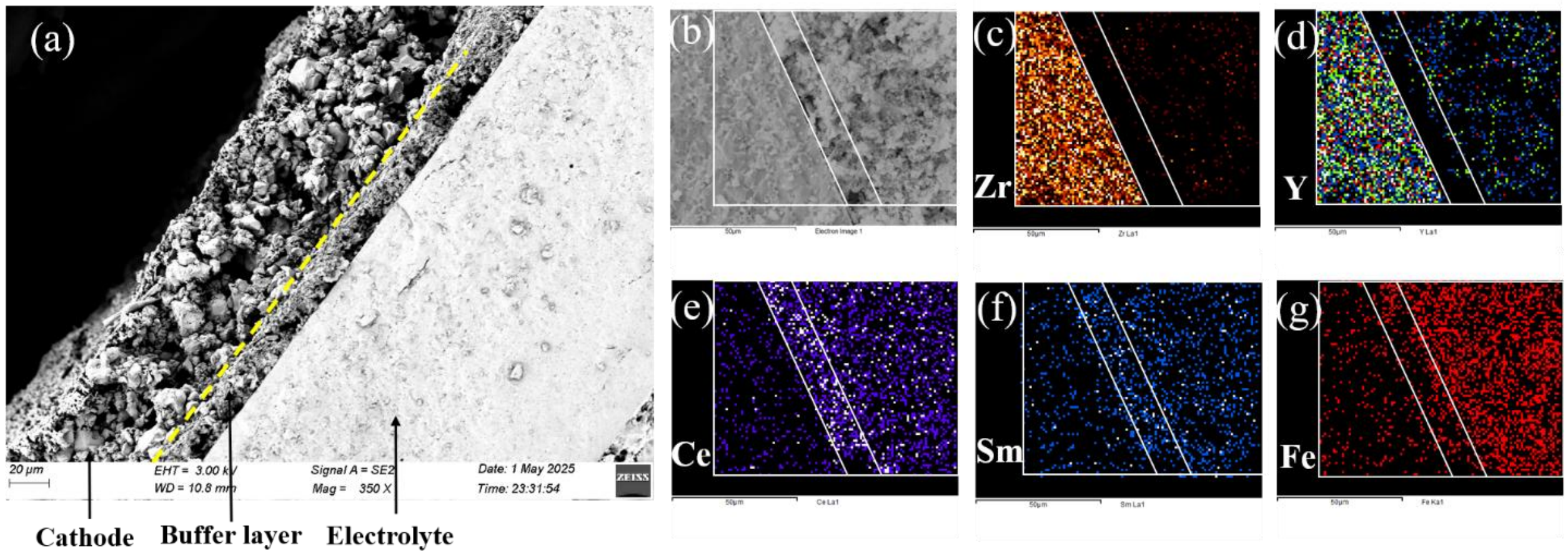

**Figure 2:** (a) Cross-sectional FESEM images of the symmetric cell with configuration STF/IF#|SDC (Buffer layer) | YSZ | SDC (Buffer layer) | IF#/STF; (b) cross-sectional cell images for EDS; EDS mapping (c) Zr; (d) Y; (c) Ce; (f) Sm; (f) Fe

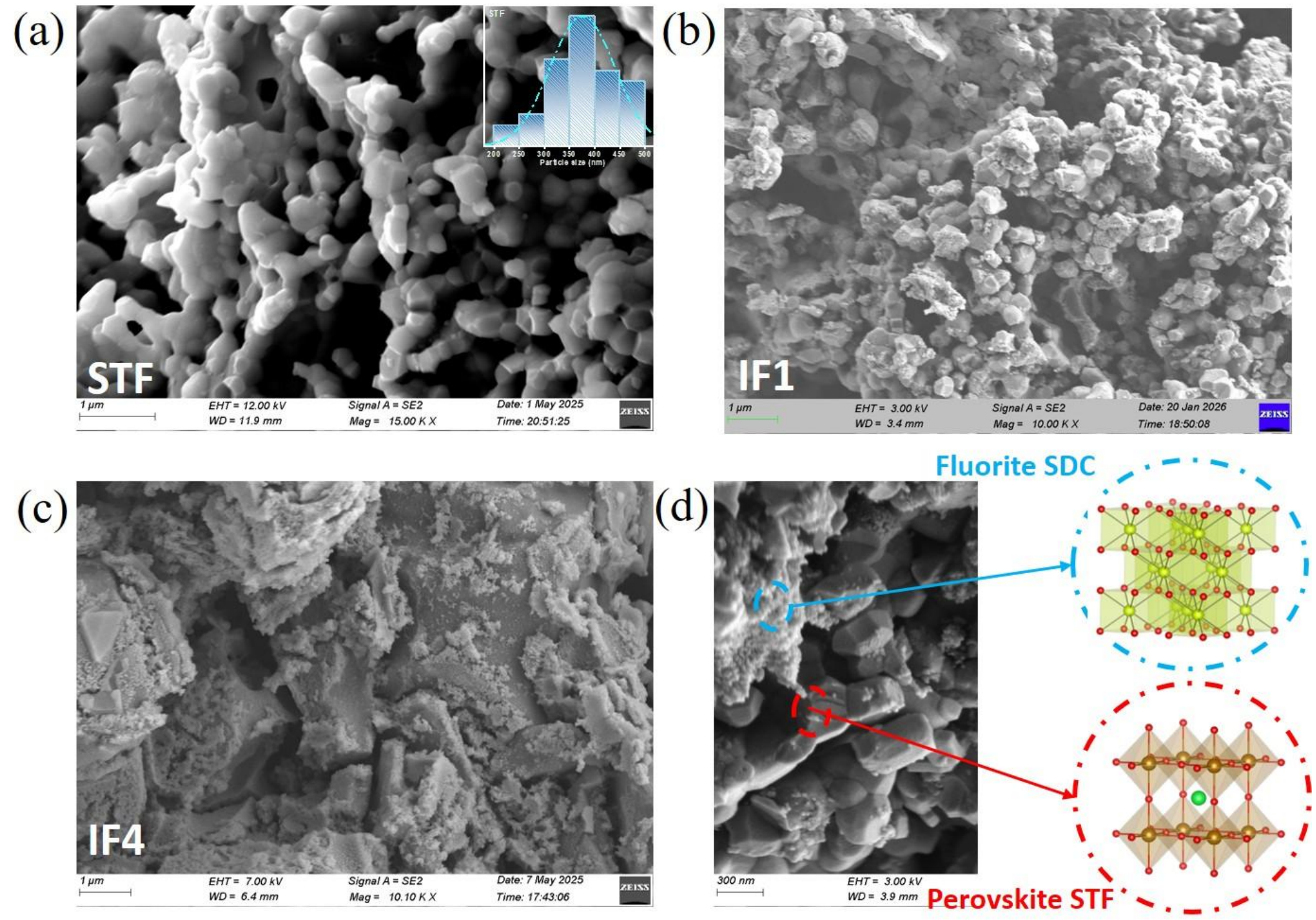

**Figure 3:** SEM images for morphology visualizations of (a) $SrFe_{0.90}Ti_{0.10}O_{3-\delta}$ (Cycle 0) (inset: average particles size distribution), (b) IF1 (Cycle 1), (c) IF4 (Cycle 4), (d) schematic of distribution of STF and SDC phase

It is expected that the IF-solution incorporation into these STF cells will encompass their surface. A subsequent heating at 600 °C is supposed to create SDC nps on the surface of the STF (Figure 4a). The XRD of this prepared cell shows a hump-like feature at the 20° - 30° region in the XRD background (Figure S3a). This feature can be associated with partially disordered SDC nps on the STF surface. Apart from this amorphous hump, prominent peaks corresponding to YSZ, SDC, and STF were observed.[31,32] In addition, no impurity peaks corresponding to non-conducting Zirconia phases like $SrZrO_3$, etc., were detected from the sintering of YSZ/SDC or SDC/STF interface. Due to the presence of multiple phases (STF, SDC, and YSZ), a reliable Rietveld refinement was challenging, and the overlapping peaks of the three phases made it even more challenging. However, the peak profiles tentatively matched the data. Hence, the peaks that represented only SDC and STF were isolated and fitted using pseudo-Voigt peak profiles. For the SDC phase, the (111) peak at ~ 28.25° is an isolated one (Figure 4b). The crystallite size was estimated to be ~ 53 nm for the symmetric half-cell before infiltration. The X-rays has penetrated through SDC buffer layer; hence, this crystallite size is typically of the SDC buffer layer. Note that the crystallite size of the sol-gel prepared SDC powder was ~ 23 nm. Upon high-temperature sintering at 1350°C, the crystallite size of the SDC buffer layer had grown to ~ 53 nm. Upon infiltration, this SDC peak revealed an extra broadening near the base. These broad peaks and their corresponding backgrounds may be responsible for the observation of a broad background in the low 2θ region. Note that any contribution from the substrate holder or other possibilities can be ruled out, as this nature is completely absent in the pristine STF cell XRD, which was measured in a similar environment. Therefore, the SDC (111) peak of the IF cell was fitted with two components of different crystallite size distributions. While one size was extremely small, ~ 10 nm, likely originating from the newly grown infiltrated SDC nps on the STF surface, the other size is larger than 53 nm, indicating a signal from the SDC buffer layer. This was visually observed in the FESEM images as well. The pores of the STF layer seem to be increasingly filled with the number of infiltrations (Figure 3b, c). On the other hand, an isolated STF peak was observed at ~ 40.20° (Figure S3b). The peak doesn't show major changes in peak position and FWHM between the pristine STF-cell and IF4 cells. Hence, the infiltration probably does not chemically modify the bulk STF phase.

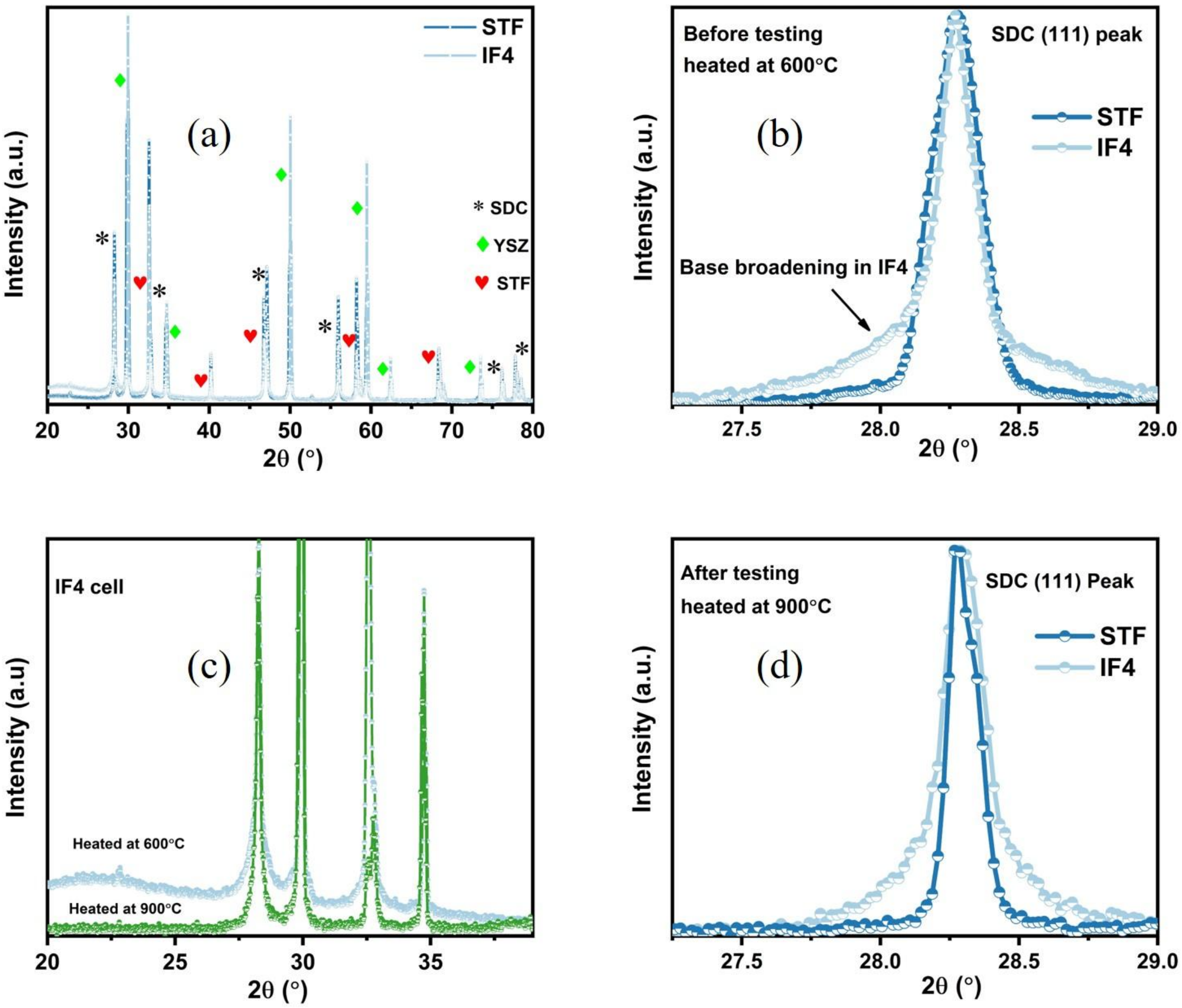

**Figure 4:** (a) XRD pattern of the STF and IF4 cell with indicator of different constituent interfaces heated at 600°C before test (Electrolyte (YSZ), Buffer (SDC) and Cathode (STF)); (b) SDC (111) peak base broadening comparison for STF and IF4 Cell heated at 600°C (c) XRD pattern of IF4 cell with before and after test (heated at 900°C), (d) SDC (111) peak base broadening comparison for STF and IF4 Cell heated at 900°C

An important observation from the XRD of 600 °C-treated IF cells was that though the multiple IF cells undergo repeated heating at 600°C, the amorphous isn't affected much (Figure S3a). However, with Pt paste baking at 900°C, the humpy background vanished when the cell XRD was taken after testing (Figure 4c). After the first infiltration and subsequent heating at 600 °C, the amorphous layer should act as a seeding layer for further growth of nps. Hence, with multiple infiltrations, the number of nps will increase, being seeded by the initial amorphous layer, thereby gradually covering the entire STF surface by successive infiltration. However, with the application of Pt paste and subsequent annealing at 900 °C for 1 hr during the testing, the amorphous layer should crystallize to form new nps or increase the size of the existing nps. The crystallite size was calculated from the SDC (111) peak from the Cell XRD of IF cells after testing (Figure 4d). It shows a narrower base broadening with crystallite size increased to 20-22 nm. This slight increase in infiltrated SDC nps shows importance of the seeding layer created due to threshold heating up to 600 °C during repeated cycles and prevents particle coarsening during thermal cycling. The FESEM images of the after-tested IF cells show increasing density of SDC nps from IF1 to IF4. From these observations, it is believed that the electrode is composed of a perforated STF matrix of interconnected STF microcrystals covered by layers of an agglomerated network of SDC nps. The masses of the samples were measured before and after each infiltration, and the mass loading percentage was calculated w.r.t. the parent STF sample. It increased from 8.2% in IF1, 16.4 % in IF2, 24.6 % in IF3 to 32.8% in IF4.

The high-resolution spectra of Fe 2p and Ti 2p were analyzed pristine STF samples (Figure 5). The features were deconvoluted after deducting a Shirley background. The oxidation states of Fe were carefully analyzed with contributions from $Fe^{3+}$ and $Fe^{4+}$ components for both the Fe $2p_{3/2}$ and Fe $2p_{1/2}$ states with a spin–orbit splitting (SOS) of ~13.4 eV, while the Ti spectra seem to be predominantly composed of contributions from $Ti^{4+}$ for Ti $2p_{3/2}$ and Ti $2p_{1/2}$ components, with a SOS of approximately 5.8 eV. [33,34]

The Ti 2p XPS spectrum of STF was fitted with $Ti^{4+}$ contributions for both the Ti $2p_{3/2}$ and $2p_{1/2}$ components without a requirement of $Ti^{3+}$ component (Figure 5a). The $Ti^{4+}$ $2p_{3/2}$ and $Ti^{4+}$ $2p_{1/2}$ features were centered at ~ 458.17 eV and ~ 463.97 eV, respectively. A weak satellite feature was observed at ~ 471.85 eV. These results indicate that Ti is predominantly present as $Ti^{4+}$ in STF. An additional feature around ~ 461 eV in the Ti $2p_{1/2}$ region is attributed to the intrinsic broadening arising from multiple unresolved final states, including normal, single-hole, and shake-up configurations, as reported in theoretical studies.[35] In the Fe-2p spectra, the $Fe^{3+}$ $2p_{3/2}$ and $Fe^{4+}2p_{3/2}$ were centered at ~ 710.72 eV and ~ 712.97 eV, respectively (Figure 5b). These values are consistent with reported values for mixed-valence iron-based perovskite oxides.[27] Quantitative analysis based on integrated peak areas reveals a mixed $Fe^{3+}/Fe^{4+}$ ratio of approximately 50.6/49.4 (effective concentration of $Fe^{3+}/Fe^{4+}$ 45.6/44.4) in STF. Hence, the ratio of $B^{3+}/B^{4+}$ ions is effectively 45.6/54.4 as compared to 43.0/57.0 for pure SFO,[25] thereby decreases the possibility of hole hopping via Fe-O-Fe pathway in STF.[27] Note that, the possibility of an oxygen vacancy in such an SFO lattice is more probable in the vicinity of a 3+ cation. This associates the electron hopping mechanism to be correlated with oxygen transport. Therefore, $Ti^{4+}$ substitution enables the increase of the presence of $B^{3+}$ ions consistent with previous reports, thereby facilitating electron exchange and as a result oxygen transport.[27,36] The peak area ratio is summarized in Table-1.

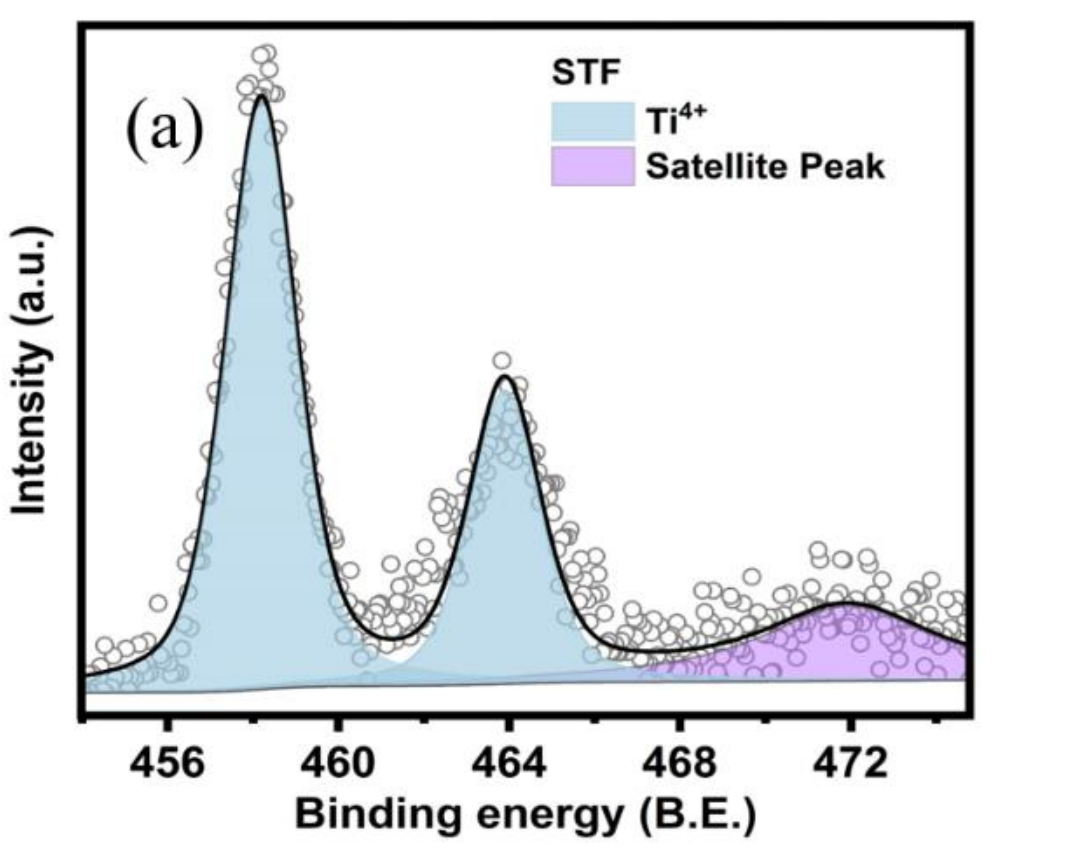

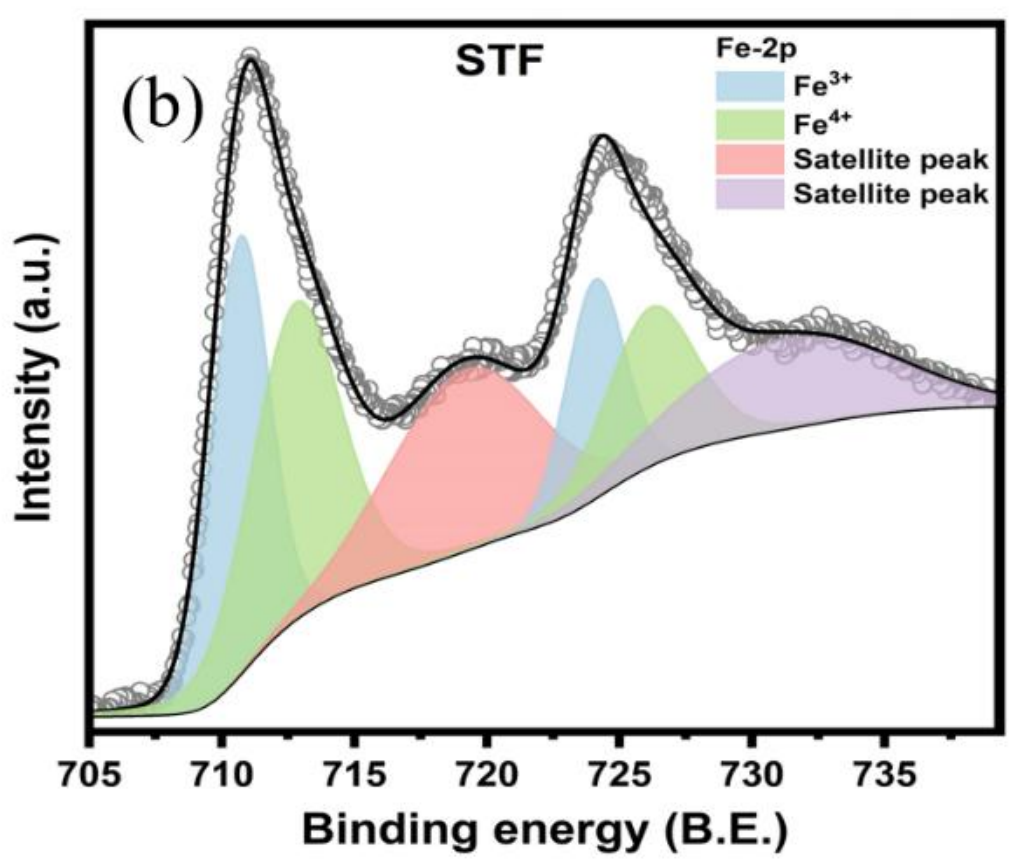

**Figure 5:** Deconvoluted XPS of $SrFe_{0.90}Ti_{0.10}O_{3-\delta}$ (a) Ti-2p (b) Fe-2p

**Table 1.** Area fraction contribution of oxidation states of constituent element of powder $SrFe_{0.90}Ti_{0.10}O_3$ obtained from XPS peak fitting

| Sample | $Fe^{3+}$ | $Fe^{4+}$ | $Ti^{4+}$ | Avg. valency of B-site cation | Oxygen deficiency (δ) |
|---|---|---|---|---|---|
| STF | 50.6 | 49.4 | 1.0 | 3.54 | 0.23 |

EIS measurements of the half-cells in the frequency range of 0.1 Hz – 2 MHz were conducted in an $O_2$ atmosphere, with 20 mV AC perturbation under open circuit conditions. Nyquist plots revealed two distinct semicircular arcs in the operating temperature range of 650°C -800°C (Figure 6a-e) A high-frequency arc is generally attributed to fast processes like charge-transfer, and a low-frequency arc is associated with comparatively slower surface-related phenomena.[18,37,38] These overlapping processes were deconvoluted using complex non-linear least squares (CNLS) fitting. The equivalent circuit employed was $R_0$ - ($R_1$||CPE1) - ($R_2$||CPE2), where $R_0$ represents an ohmic resistance, $R_1$ and $R_2$ correspond to the high and low-frequency contributions of resistance, respectively, while CPE1 and CPE2 are constant phase elements (CPEs) that were introduced to account for non-ideal interfacial capacitive behavior matching the high and low frequency regimes. The polarization resistance ($R_P$) is defined as $R_P = R_1 + R_2$, and the area-specific resistance (ASR) is expressed as ASR = ½ $R_P$. The factor ½ is due to the similar electrodes on both sides. The fitted arcs are shown for STF and IF3 at 750°C in figure - 6f.

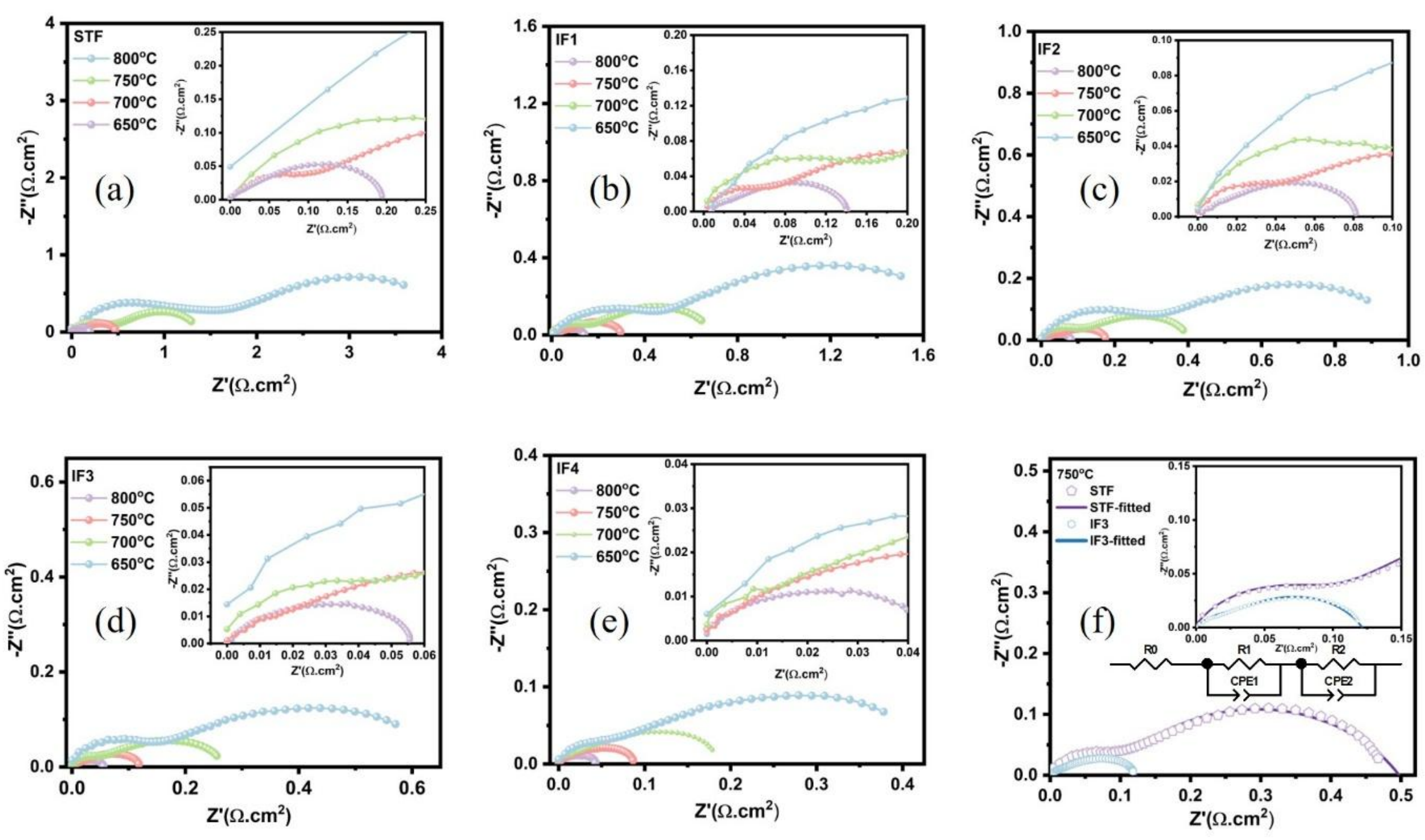

**Figure 6:** Nyquist plot comparison of various symmetric pure and infiltrated cells data taken at different temperature (a) $SrFe_{0.90}Ti_{0.10}O_{3-\delta}$ (Cycle 0) (b) IF1 (Cycle 1) (c) IF2 (Cycle 2) (d) IF3 (Cycle 3) (e) IF4 (Cycle 4) (f) equivalent circuit fitted for STF and IF3

For the pristine and IF samples, these components, $R_0$, $R_1$, and $R_2$, reduce nearly exponentially with temperature from 650°C to 800°C Ω·cm² for all the samples (Figure S4, 7a, 7b). The rate of decrease is maximum for the pristine STF, decreasing from 4.05 Ω·cm² to 0.18 Ω·cm² (~26 times less). For the infiltrated samples, the rate of decrease with temperature is reduced but remains considerable (~ 10 times lesser for IF1 and ~ 13 times lesser for higher infiltrations). The comparison is shown in Table 2. Note that with the increasing infiltration, the resistances also decreased nearly exponentially at a particular operational temperature (Figure 7c, 7d). A comparison of various contemporary air electrode's ASR values with the SDC infiltrated STF has been tabulated in Table – S2.

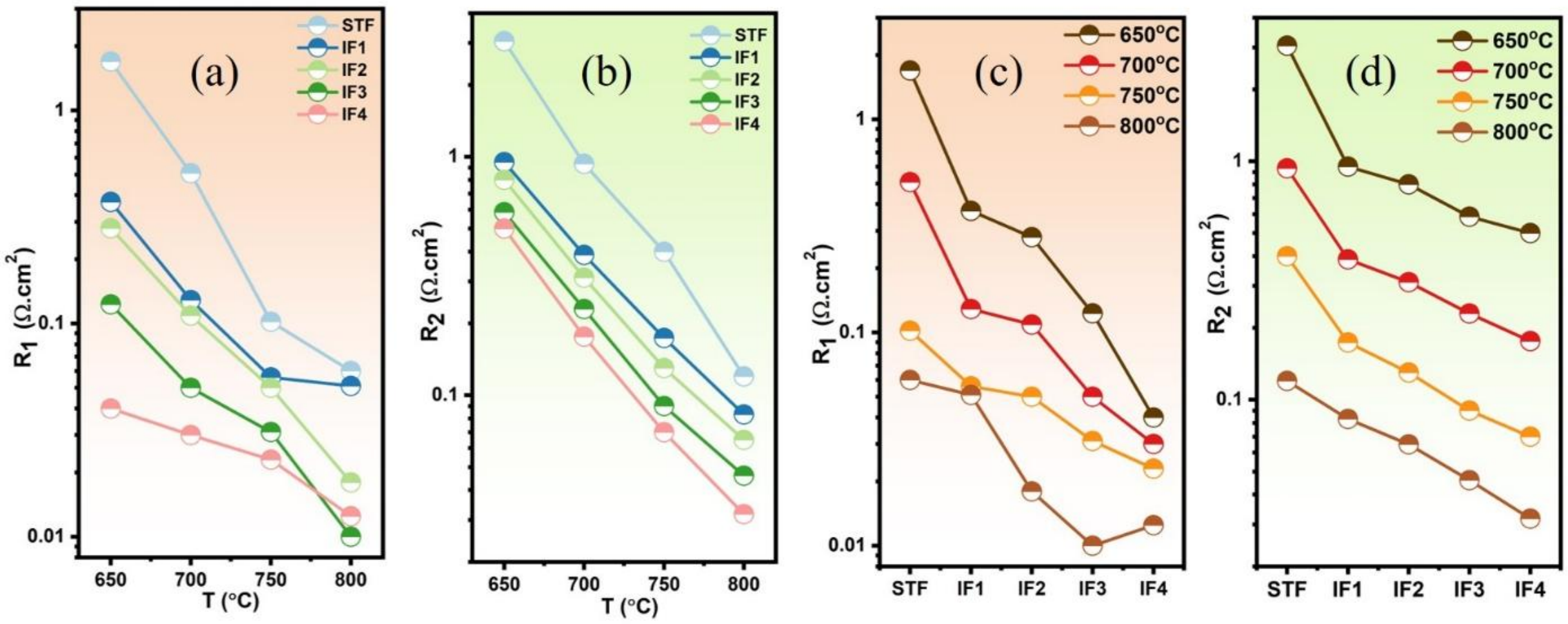

**Figure 7:** Symmetric pure and infiltrated cells EIS circuit fitting parameter (a) $R_1$ at different temperature (b) $R_2$ at different temperature (c) $R_1$ at different composition (d) $R_2$ at different composition

**Table 2.** Temperature dependent polarization resistance ($R_P$) ($\Omega.cm^2$) for STF and infiltrated cells obtained from CNLS fitting of impedance spectra

| Temperature (°C) | STF | IF1 | IF2 | IF3 | IF4 |
|---|---|---|---|---|---|
| 650 | 3.94 | 1.24 | 0.96 | 0.59 | 0.41 |
| 700 | 1.29 | 0.48 | 0.37 | 0.25 | 0.18 |
| 750 | 0.46 | 0.21 | 0.16 | 0.11 | 0.09 |
| 800 | 0.16 | 0.11 | 0.07 | 0.05 | 0.04 |

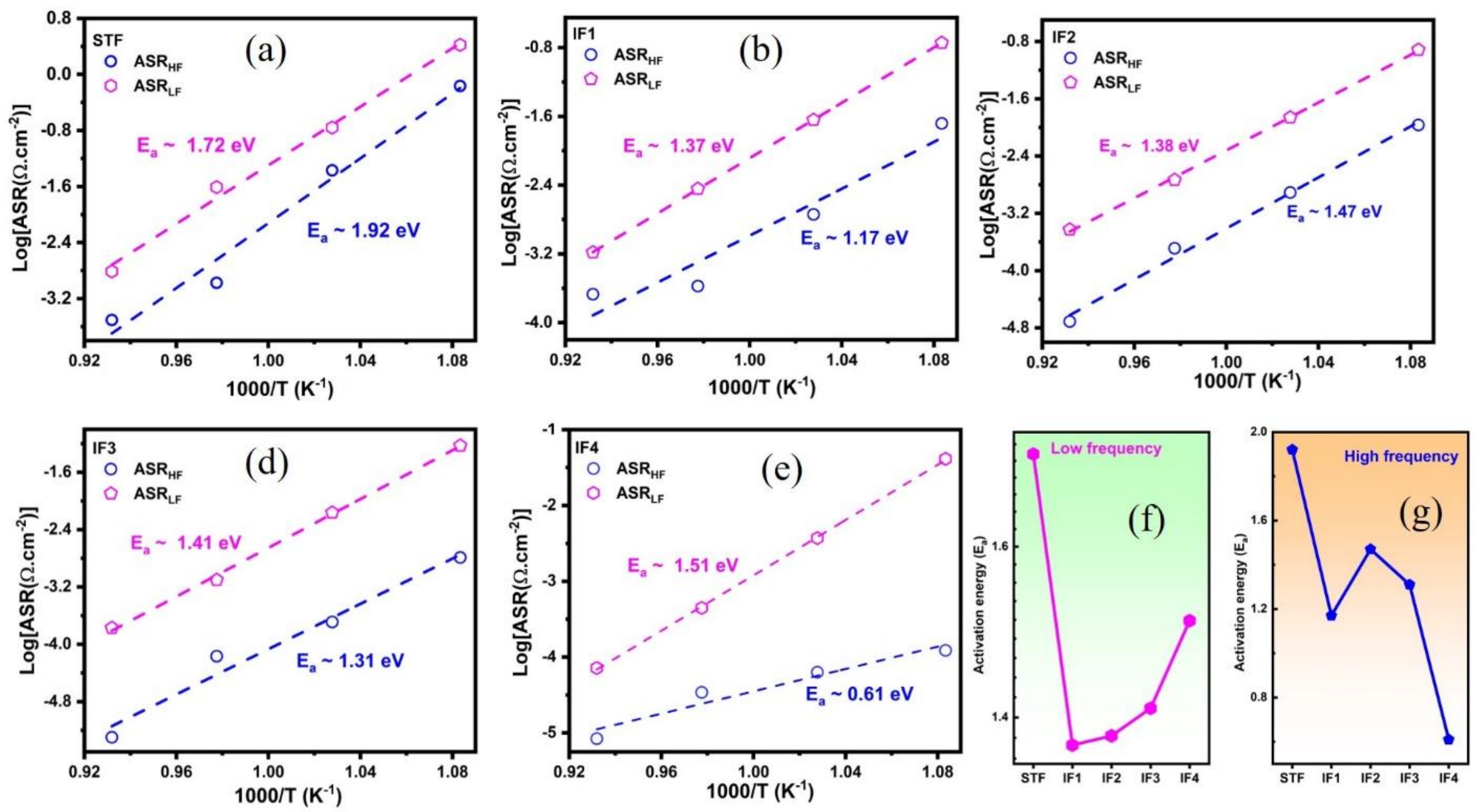

**Figure 8:** Activation energy (E$_a$) calculation for (a) STF, (b) IF1, (c) IF2, (d) IF3, (d) IF4, evaluation of Ea of (f) Low (green) and (f) High (orange)

To gain an insight into the nature of the EC processes, Ea was calculated from the slope of the linear plots of log (ASR) vs. 1000/T (Fig-8a-e) for separately LF and HF processes. A deviation from linearity is evident in these plots, implying a deviation from a perfect Arrhenius nature ($R = R_0 \exp(\frac{-E_a}{RT})$, where E$_a$ is the activation energy of the process involved. The deviation is more evident in HF fitting, i.e., for R$_1$. This suggests that there is more than one activation process involved in these materials. For the LF process, the activation energy (E$_a$) decreases drastically from 1.72 eV in STF to 1.37 eV in IF1, followed by an exponential rise in IF2 (1.38 eV), IF3 (1.41 eV), and IF4 (1.51 eV) (Figure 8f). This rise indicates an internal shift in the rate-determining steps during surface redox activity, driven by infiltration on the slow time scale. However, to decipher which process is exactly contributing to the rate-determining step, a detailed study of DRT is required, which is especially capable of deconvoluting the overlapping processes. On the other hand, in the HF process, E$_a$ decreases substantially from 1.92 eV (STF) to 1.17 eV (IF1), 1.47 eV (IF2), 1.31 eV (IF4), and 0.61 (IF4) (Figure 8g). Though E$_a$ of IF cells is considerably less than STF, the value doesn't have continuous change. Hence, the underlying processes and their overlapping nature will be discussed in the DRT section in detail. Overall, the high activation energy eliminates the possibility of gas diffusion related polarization resistance.[39]

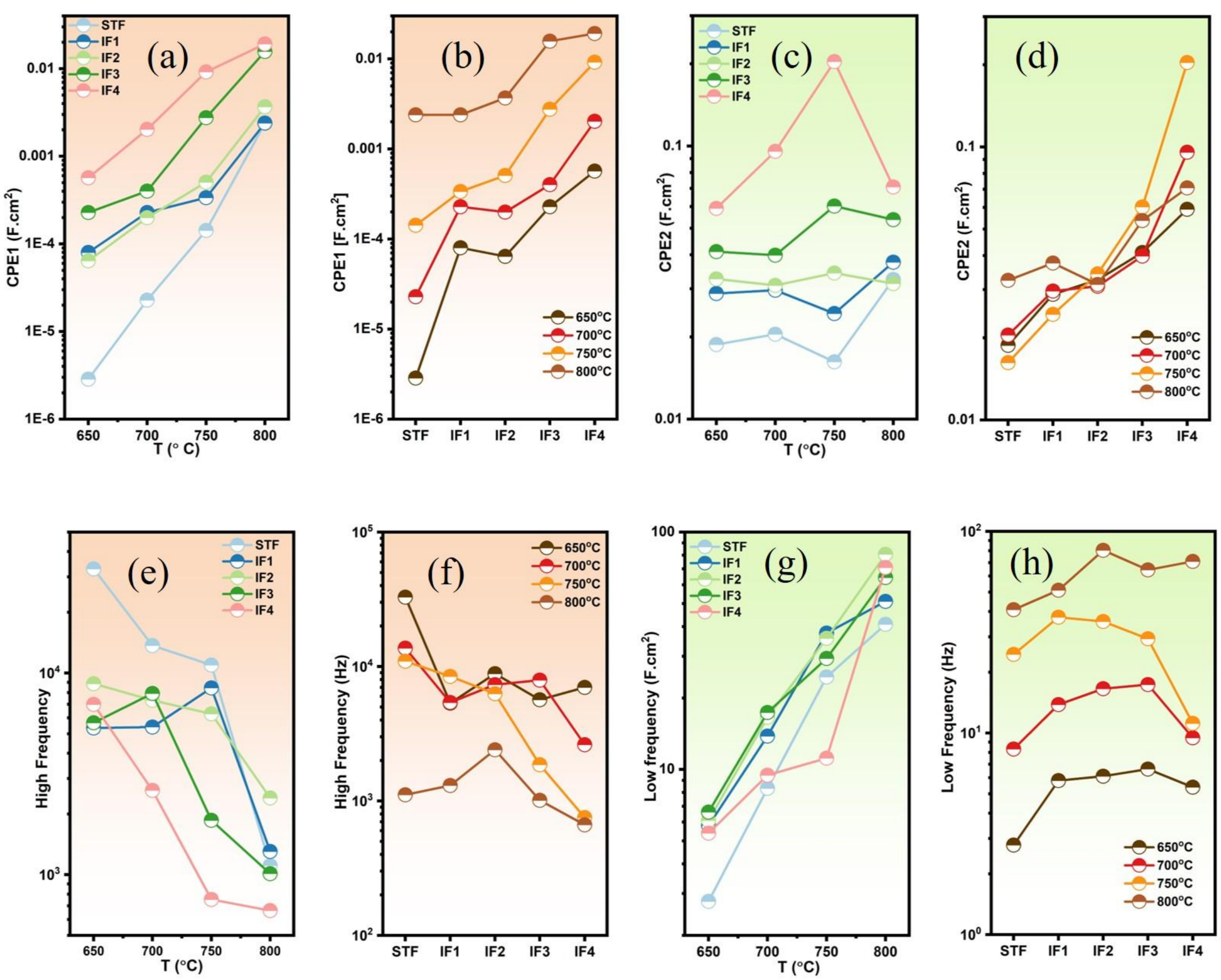

**Figure 9:** Circuit fitting parameter comparison of different cell (a) CPE1 with temperature; (b) CPE1 with composition; (c) CPE2 with temperature; (d) CPE2 with composition; (e) high frequency with temperature; (f) high frequency with composition; (g) low frequency with temperature; (h) low frequency with composition;

On the other hand, the capacitance for the high (CPE1 or C1) and low (CPE2 or C2) frequency processes was also calculated using the formula ($C = Q^{(1/n)} R^{(1-n)/n}$), where Q = amplitude of the capacitance and n = phase fraction of the capacitance. While C1 increases with temperature and infiltration concentration (Figure 9a, b), C2 remains moderately invariant with temperature except for IF4, which increases with infiltration at all temperatures (Figure 9c, d). Note that the changes by a few orders in C1 with both temperature and infiltration, while the changes in C2 are hardly an order different. The corresponding resonant frequency associated with each circuit was also calculated using the formula $f = (1/2\Pi RC)$, where (R, C) represents either (R1, C1) or (R2, C2) for the HF and LF regimes. It was observed that the HF processes slow down by an order of magnitude (from ~ 10kHz to ~ 1kHz) with the increase of temperature (Figure 9e). With increasing infiltration, the HF process shifts towards lower frequency, despite minor fluctuations at intermediate infiltration level (Figure 9f). At 800 °C, the speeds are comparable though with IF2 being the fastest. On the other hand, the LF processes become faster by an order of magnitude from ~10 Hz to ~100 Hz with temperature (Figure 9g). Increasing infiltration up to IF3 results in a systematic increase in the LF characteristic frequency (Figure 9h). Although the IF4 sample exhibits lower characteristic frequencies than the other infiltrated samples at lower temperatures, its frequency becomes comparable to the others at 800 °C. Similar to the activation energy analysis, identifying the underlying processes is essential to explain the observed changes in their characteristic frequencies.

The temperature-dependent EIS data were further analyzed through distribution of relaxation times (DRT) spectra (Figure 10a-e) to gain better insight of the overlapping EC processes. These spectra were deconvoluted using Gaussian functions, and the integrated area of each peak was recorded as the resistive contribution of their respective process. The evolution of peak resistance and characteristic frequency with temperature is summarized in figure 11a, b. Eight relaxation processes (P1–P8) were resolved in the frequency range of approximately 3 Hz to 1 MHz, suggesting they are contributing towards the rate limiting steps. All the features decreased in terms of resistance and revealed systematic shifts in the characteristic frequency with increasing temperature.

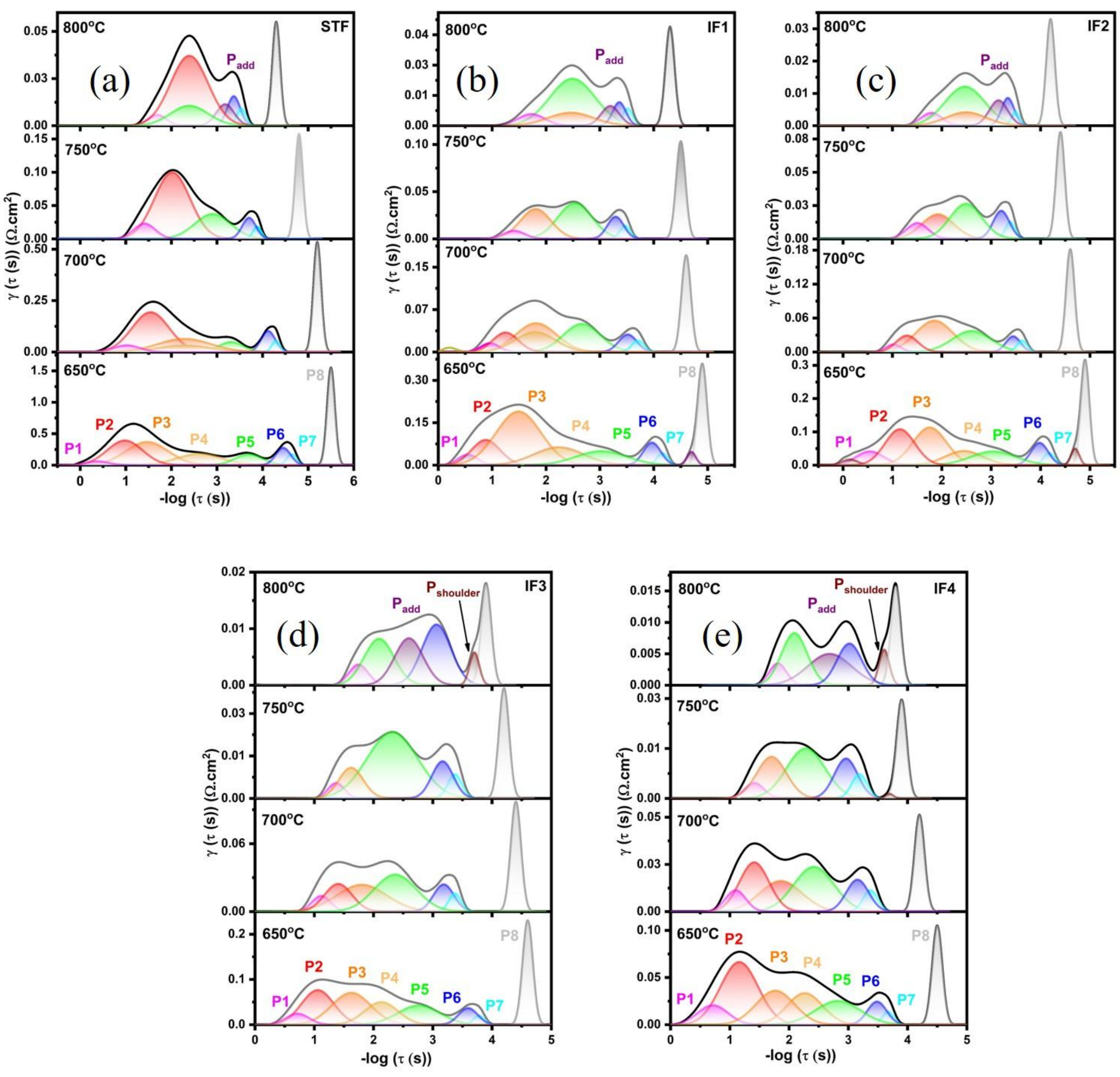

**Figure 10:** Deconvoluted peaks of Distribution of relaxation time (DRT) spectra of symmetric cell EIS data of (a) STF, (b) IF1, (c) IF2, (d) IF3, (e) IF4

As the oxygen molecules approach the surface of the cathode a Van-der-Walls force acts between the active site of the cathode and the $O_2$ – molecule in order to reduce the thermodynamical energy. This process leads to the physio-adsorption of the $O_2$ – molecule on the surface and are generally slower processes. At this stage an electron from the active site initiates the chemi-adsorption process of dissociation of adsorbed $O_{2,g}$ molecule ($O_{2,ads}$) to form two adsorbed O atoms ($2O_{ads}$) by breaking the π-bond. Hereafter, the processes of charge transfer is initiated. The 1$^{st}$ charge transfer adds a single electron to the $O_{ads}$ to form $O_{ads}^{-}$, followed by a 2nd charge transfer forming $O^{2-}$ ions. The $O_{ads}^{-}$ , and $O^{2-}$ are free to migrate on the surface and after transforming to $O^{2-}$ at the active TPB sites, diffuse into bulk as ionic transport carriers. Hence, the overall pathway in the MIEC cathodes can be described as: $O_{2,g} \rightarrow O_{2,ads} \rightarrow 2O_{ads} \rightarrow O_{ads}^{-} \rightarrow O^{2-}$.[18,37,40,41]

The above transformation of $O_{2,\,g}$ molecule to a diffused $O^{2-}$ ions involves transformation pertaining to different speeds of relaxation. Based on this, these processes can be identified with different characteristic frequencies. These are sensitive to temperature and the nature of the active sites on the surface. Hence, the DRT spectra in the present study is a powerful tool in understanding each of these processes and can be divided into low-frequency (LF), medium-frequency (MF), and high-frequency (HF) regimes. P1 and P2 in the LF (~ $10^{-1}$ - $10^{2}$ Hz) region are assigned to slower surface oxygen exchange processes. P3–P5 in the MF (~ $10^{2}$ - $10^{4}$ Hz) region are associated with intermediate oxygen reduction pathways associated with oxygen activation and near-surface electrochemical reactions. P6 and P7 in the HF (~ > $10^{4}$ Hz) region are attributed to rapid oxide-ion transfer near the electrode/electrolyte interface.P8 is the highest-frequency contribution and is mainly associated with the ohmic response of the cell. [4,41–43] The proposed assignments are further supported by the infiltration results. Since SDC is known to have high $V_O$ concentration,[44] relative changes in DRT peaks following infiltration provide insight into the extent to which the corresponding processes are coupled to surface defect density.

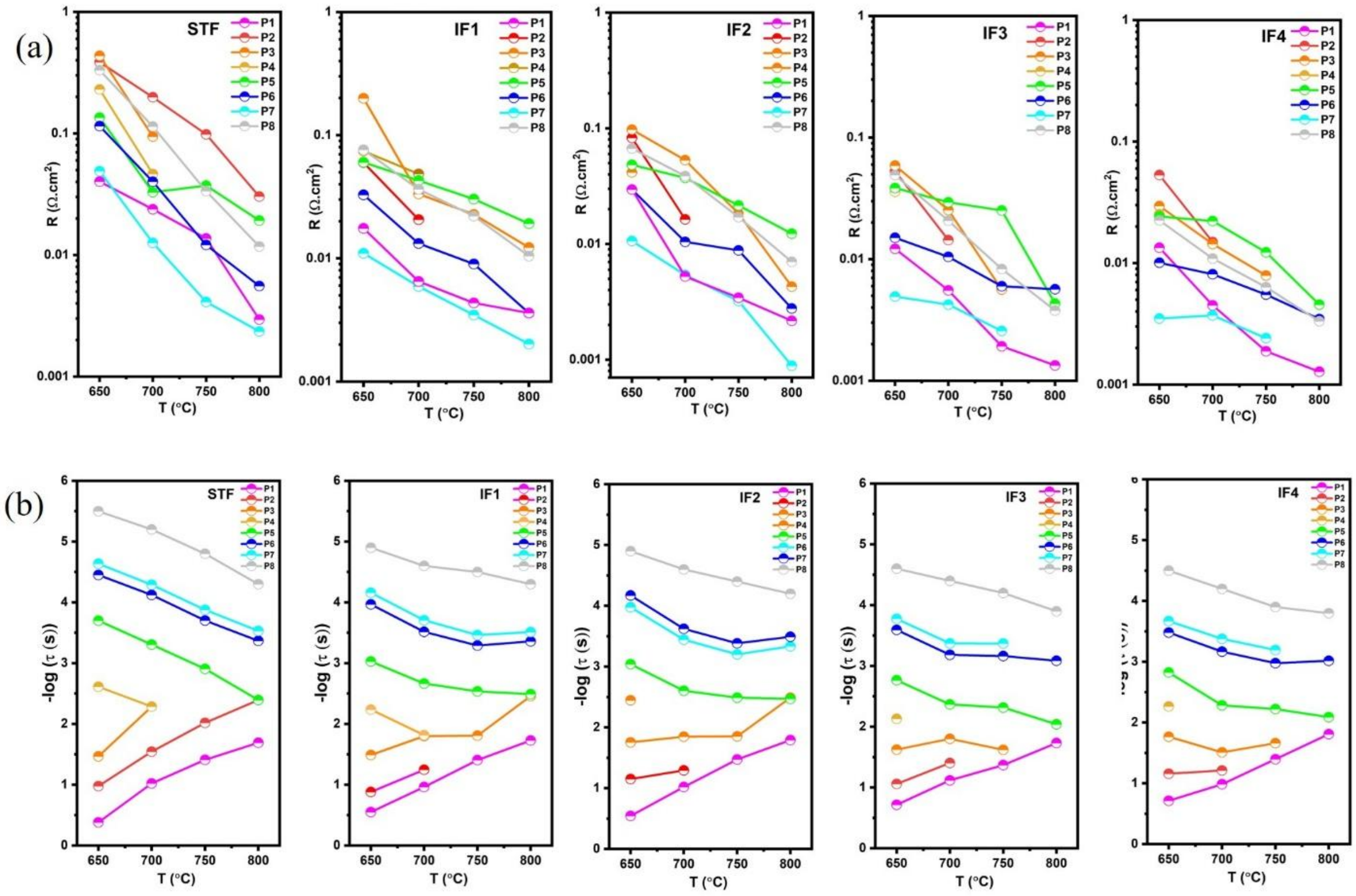

**Figure 11:** DRT fitted peak parameter evolution with temperature for different cells (a) Peak area (resistance) (b) Peak center (frequency)

Pristine STF: The DRT LF spectrum at 650 °C for pristine STF was composed of a prominent P2 and a weak P1. This indicates that the surface oxygen exchange was difficult for the STF sample at the lowest temperature. The resistance related to this process reduced with temperature and moved towards higher frequencies (Figure 12a, b). Such a result can be justified as surface oxygen exchange like adsorption/desorption and onset of the early dissociation steps becoming faster and easier with increasing temperature. The nature of the P1 process resembles that of P2. Note that P1 is a similar process of surface oxygen exchange, probably with a different origin due to heterogeneity of adsorption sites, but with much less resistance, i.e., a more prominent process. Note that the oxygen diffusion process is also an LF process, but is generally accepted as being temperature-insensitive.[45] Hence, P1 and P2 being temperature dependent should not be attributed to an oxygen diffusion process. However, one can argue that as the temperature increases, the rising kinetic energy of the gas particles may facilitate faster diffusion through the porous electrode structure. This should reduce the gas diffusion resistance and improve overall cell performance. However, even at 650 °C, any such feature is not observed. Further, even at high temperatures, there is no observation of such a feature for STF. Hence, for the STF samples, LF processes are attributed primarily to surface-controlled oxygen exchange rather than gas diffusion limitations.

Amongst the MF processes, P3, P4, and P5 are the prominent ones. They are broadly associated with oxygen activation and intermediate reduction kinetics ($2O_{ads} \rightarrow O_{ads}^{-} \rightarrow O^{2-}$) occurring at and near the electrode surface. The slowest P3 process can be correlated to oxygen dissociation and subsequent electron uptake to form oxygen intermediates ($O^{-}/O^{2-}$). The feature shows a reduction in resistance with temperature and eventually merges with P4 at 700 °C, and thereafter both vanish at 750 °C and above. Note that the P4 process is faster than P3 and can be attributed to similar electrochemical activity of adsorbed oxygen, a subsequent step of P3. Its separate existence at low temperature could be due to active site heterogeneity. As the temperature rises due to an increase in molecular dynamics, P3 merges with P4 and can no longer be individualized.

The effect of infiltration has a major effect on the surface kinetics. Although the P1 process doesn't become drastically slower or quicker, the resistance of P1 seems to decrease at all temperatures with the number of infiltrations (Figure 12a magenta). P2 shows a similar behavior in terms of speed except the process is not a rate limiting step anymore above 700°C for all the infiltrated samples. The resistance shows a major reduction after the 1st cycle, followed by a gradual decrease with further mass loading (Figure 12a red). This indicates an easier and faster surface oxygen adsorption with increasing infiltration and temperature. At 650°C, for IF1 the major contribution in the DRT comes from P3 while for the rest of the cells it is from P2. In all the IF cells P3 merges with P4 at 700°C similar to pristine STF which may stem from enhanced oxygen spill-over effect for adsorbed oxygen intermediates at elevated temperatures.[46,47] Another important observation is that with increasing infiltration P3 decreases exponentially i.e., surface oxygen activation and reduction become easier as SDC increases the active TPB site. It is important to note that the process not only becomes easier but also becomes faster with temperature but slows down with infiltration.

Moving on to the faster processes in the MF regime, resistance of P5 in STF show slight deviation from the Arrhenius behavior ($E_a$ ~ 0.52 eV) (Figure-S5) in STF. P5 shows gradual decrease in resistance with temperature in all IF cells. In IF1 and 2 the P3 process merges with P5 at 800°C with much smaller resistance. Moreover, the relative change in P5 resistance upon infiltration is not as drastic as compared to other processes (Figure 12a green). Consequently, it emerges as the dominant feature in DRT spectra of the infiltrated electrodes at temperatures above 700°C. In addition, the frequency of the processes decreases with both temperature and infiltration (Figure 12b-green). This observation suggests the origin of the process is coupled with STF surface which could be electron transfer from STF to surface active O-molecule. At 800°C, an additional contribution ($P_{add}$) appears adjacent to P5 on the higher-frequency side most probably previously embedded in to P5. Its resistance too is not much effected with infiltration (Figure-S6).

In the HF regime, P6 and P7 peaks are very closely spaced in terms of frequency and become slower with both temperature and mass loading. They follow the Arrhenius behavior strongly for all the cells indicating the associated EC process are strong thermally activated in nature. Ea value changes from 0.92 eV in STF to 0.37 eV in high infiltrated samples (Figure-S5). The peaks can be associated with surface oxygen incorporation ($O^{2-} + V_O \rightarrow O_O^x$) and local oxide-ion transport though bulk. Similar to P2, a drastic decrease in the resistance from STF to IF1 is observed at 650 °C. The magnitude of this improvement decreases at higher temperatures, and by 800 °C, the resistances of all samples are low and comparable. Further infiltration leads to only nominal reductions in resistance, suggesting that these interfacial processes approach a saturation limit once a sufficiently connected ionic network is established. With more TPB site generation with infiltration and decrease in activation barrier for oxide-ion formation and migration, oxygen ions are generated and accommodated more readily at the surface and at STF–SDC interfaces. If the rate of oxygen incorporation and transport does not increase proportionally, a larger concentration of mobile oxide ions and oxygen vacancies accumulates near the electrode/electrolyte interface. This increases the defect-storage capacity of the electrode and results in a longer effective relaxation time. The hint could be observed from increase HF C2 capacitance both with temperature and infiltration from CNLS fitting. A separated peak P8, could be fitted at the extreme HF regime, with Ea ~ 0.99 eV in STF (Figure-S5). The process can be associated with ohmic resistance, mainly from bulk oxide ion conduction through the electrolyte, together with minor contributions from current collectors. The resistance and frequency evolution with temperature and infiltration of this peak are too similar to P6.

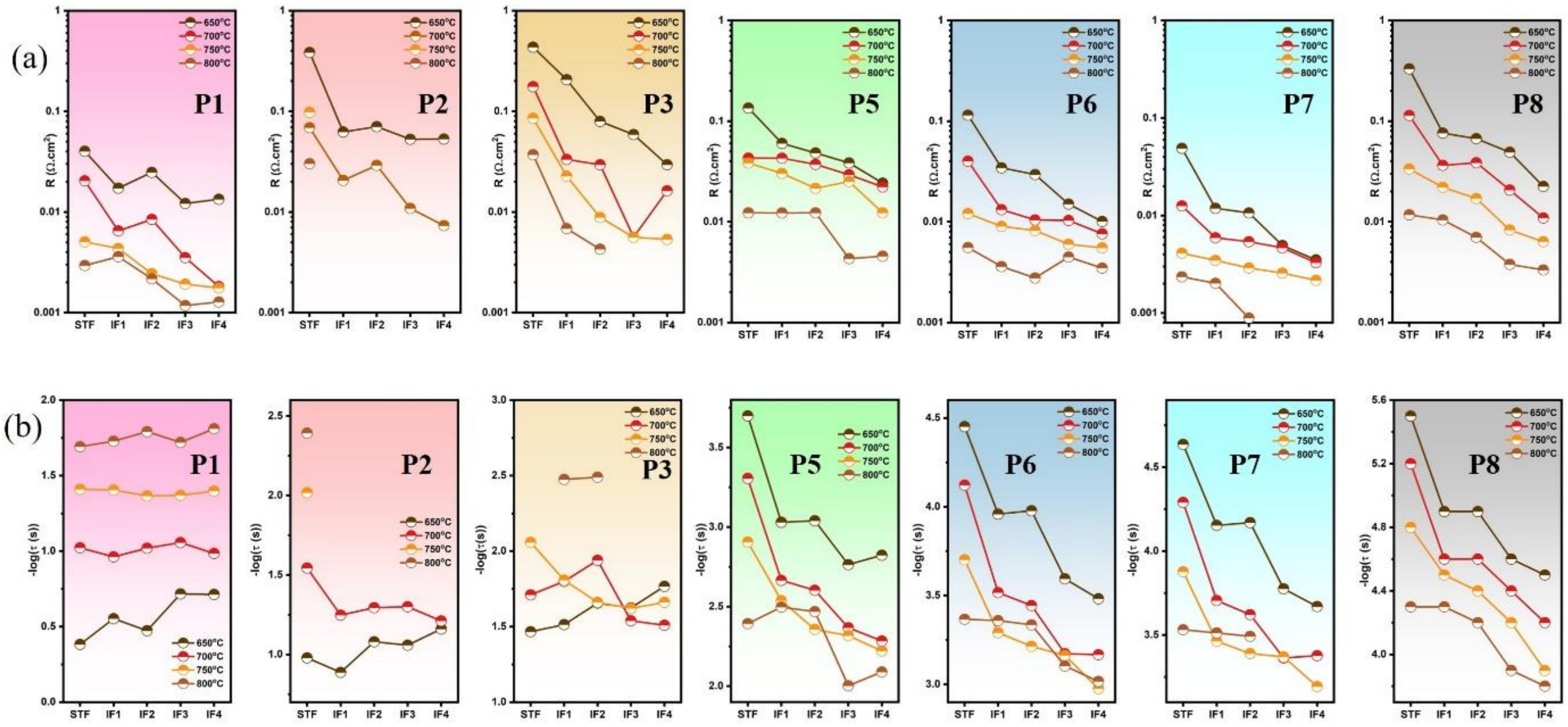

**Figure 12:** DRT fitted each peak parameter evolution with composition (a) Peak area (resistance) (b) Peak center (frequency)

At higher SDC loadings, a weak LF peak near $10^{-1}$ Hz and a shoulder adjacent to P8 become visible (Figure-S7). The LF contribution is attributed to gas diffusion and concentration polarization. Its appearance suggests that excessive SDC begins to partially block the porous network. Although IF4 exhibits a slightly lower total polarization resistance than IF-3, the difference is small. One of the most important observations is the shift in the dominant resistive contribution. In pristine STF, the main resistance above 700 °C arises from P2 and P3, associated with surface oxygen exchange and activation. After SDC infiltration, the dominant contribution shifts to P5. This indicates that SDC removes the kinetic bottleneck associated with the early stages of the ORR, leaving electron transfer to surface active O-molecule as the principal rate-limiting step.

Taken together, the major performance gains are achieved between IF3 and IF4. Beyond this point, additional infiltration produces only marginal improvements and introduces diffusion-related hindrances. The optimal loading therefore lies between IF3 and IF4, corresponding to approximately 25–33 wt.% SDC.

## Conclusion:

In summary, the present work systematically investigates the influence of SDC infiltration of 10% Ti-doped $SrFeO_{3-\delta}$ in symmetric cell configuration for the improved ORR process in SOFC air electrode. The cell XRD revealed structural stability of multiple stacked layers before and after testing, while changes in the SDC (111) diffraction peak indicated the evolution of the infiltrated nanocrystalline phase. Equivalent R-C circuit fitting of the symmetric cell EIS showed decreased $R_P$ and increased capacitance for all the IF cells at each temperature of operation which indicates enhanced $V_O$ mediated charge storage and oxide-ion transport. The obtained ASR value showed a decrease from 0.08 $\Omega.cm^2$ to 0.025 and 0.021 $\Omega.cm^2$ at 800°C in presence of $O_2$ flow for STF, IF3 and IF4 respectively.

Detailed DRT analysis enabled the resolution of multiple ORR-related relaxation processes and gain insight into their evolution with temperature and infiltration level. The results demonstrate that SDC infiltration preferentially accelerates surface oxygen exchange, activation and early reduction processes. This leads to a shift in the dominant kinetic limitation toward charge transfer related to STF surface at elevated temperatures for the IF samples. This behavior indicates that the STF–SDC heterointerface effectively expands the electrochemically active reaction zone and improves local ionic transport pathways. The appearance of additional LF features at the highest infiltration level further suggests the onset of mass-transport limitations associated with excessive SDC loading. Thus, an optimal SDC content of approximately 25–30 wt.% was identified, providing the best balance between enhanced ORR kinetics and preservation of mass transport. These findings establish a mechanistic framework for tailoring mixed-conducting perovskite cathodes through surface engineering and provide guidance for the development of high-performance STF–SDC composite electrodes.

## Acknowledgements:

Authors, RS would like to thank the Ministry of Education, Government of India, for the Prime Minister Research (PMRF ID-2102739) fellowship, and IIT Indore International Relations for providing the fellowship during the visit to South Korea under the short-term collaborative research program. This work was supported by the National Research Foundation of Korea (NRF) grant funded by the Korea government (Ministry of Science and ICT(MSIT)) (No. RS-2023-00236572). Author MK acknowledges CSIR-UGC (ID: 211610074937) for providing the fellowship. A. Mekki, K. Harrabi, and S. Sen acknowledge the support of the King Fahd University of Petroleum and Minerals, Saudi Arabia, under Grant No. DF191055 DSR project.